\documentstyle[12pt]{article}

\makeatletter
  \renewcommand{\theequation}{%
     \thesection.\arabic{equation}}
  \@addtoreset{equation}{section}
\makeatother

\def\title#1{\begin{center}{\Large\bf #1}\end{center}}
\def\author#1{\vskip 5mm \begin{center}{#1}\end{center}}
\def\address#1{\begin{center}{\it #1}\end{center}}
\begin{document}

\vspace*{-2cm}
\begin{flushright}{}\end{flushright}\vspace{-9mm}
\begin{flushright}{}\end{flushright}
\vspace{0.5cm}

\title{Quantization 
via Star Products}
\author{Takayuki HORI\footnote{E-mail: hori@main.teikyo-u.ac.jp}and ~Takao KOIKAWA\footnote{E-mail: koikawa@otsuma.ac.jp}}
\vspace{1cm}
\address{
	$^{1}$ Department of Economics, Teikyo University,\\
	Hachioji 192-0395,~Janan\\
        $^{2}$ School of Social Information Studies,\\
        Otsuma Women's University\\
        Tama 206-0035,~Japan\\ 
}
\vspace{1cm}

We study quantization via star products. We investigate a quantization scheme in which 
a quantum theory is described entirely in terms of the function space without  
reference to a Hilbert space, unlike the formulation employing  the Wigner functions.
The associative law plays an essential role in excluding the unwanted 
solutions to the stargen-value equation. 
This is demonstrated  explicitly with the $D$-dimensional harmonic oscillator.

\newpage
  

\section{Introduction}

In 1949, Moyal discovered an interesting description of  quantum mechanics,\cite{rf:1} 
called Moyal quantization, in which the Wigner function\cite{rf:2} is used as the 
phasespace distribution function, similar to that used in  statistical mechanics.  
(See Ref.\cite{rf:3} for an early work on quantization in phasespace.)  
The transformation function connecting the distribution functions at different times 
satisfies a differential equation similar to the classical equations of 
motion, where the Poisson bracket is replaced by the  Moyal bracket.

After several decades, an abstract scheme of the Moyal quantization called deformation 
quantization  was proposed.\cite{rf:4} 
This scheme is based on the general star product 
satisfying some axioms and on a deformation of 
the symplectic structure of the phasespace in the classical theory,  
and it  is formulated without a Hilbert space.
Because one gets the Poisson bracket  in the limit  $\hbar \rightarrow 0$, this 
approach makes it possible to understand the connection between the 
classical and quantum theories from  a new perspective.
This scheme is believed  to be equivalent to  ordinary 
quantum mechanics and seems to effectively include functions 
that are written in the form of the (off-diagonal) Wigner functions or  
their linear combinations.
There has been much progress in this field,  especially with regard to its mathematical 
aspects.\cite{rf:5,rf:6,rf:7}

In the present paper we set up a general framework for quantization, which is 
slightly different from the deformation quantization. 
In this formulation, the state function is not necessarily the 
Wigner function that is constructed from vectors 
of the Hilbert space. We construct this scheme without resort to 
the Hilbert space.
In our scheme, the associativity of the star product is used in defining the 
basic function space. This contrasts  with the situation 
regarding the deformation quantization, 
where the associativity is a guiding principle in deforming  the classical 
symplectic structure of the phasespace.
In our scheme, the functions describing physical states do not necessarily 
have counterparts in  ordinary quantum mechanics.
For example,  unwanted states, such as those with negative energies, can  appear 
if we do not impose the associativity of the star product. 
These state functions are excluded by  associativity. 
For this reason, our theory is conjectured  to be 
equivalent to  quantum mechanics.
This equivalence is  demonstrated explicitly for 
the $D$-dimensional harmonic oscillator.

\vspace{.5cm}


\section{Assumptions and function space}

Let us consider a  dynamical system described by real coordinates, $q_i (i=1,\cdots, N)$, 
and their canonical momenta, $p_i$.
We make the following assumptions (1)-(6).

\noindent (1) There exist a function space ${\cal S}$ of complex valued and 
integrable functions of $(q, p)$ and a  map $\star :{\cal S}\times{\cal S} \rightarrow {\cal S}$ 
satisfying the distributive and associative laws,  the Hermiticity property, 
$(f_1\star f_2)^* = f^*_2\star f^*_1$ for $f_i \in {\cal S}$, 
and the property $1\star f = f\star 1 = f$ for $f \in {\cal S}$. 
We postulate that the pair $(\star, {\cal S})$ consisting of this operator
 (which refer to as the star product hereafter) and this function space is given, 
and physical states and observables are described by functions in ${\cal S}$.

\noindent (2) There exists a set of observables, ${\cal O}=\{O_1, O_2,\cdots, O_r\}$, 
that are mutually commuting with respect to the star product. 
Here $r$ may or may not be finite.

We  can define the eigenvalue problem associated with the star product as in the 
theory of vector spaces.
For  functions $g^i \in {\cal S}$ the function $f_{\{\lambda\}\{\lambda'\}}\in{\cal S}$ satisfying
\begin{eqnarray}
          g^i\star f_{\{\lambda\}\{\lambda'\}} &=& \lambda^if_{{\{\lambda\}\{\lambda'\}}},\\
          f_{\{\lambda\}\{\lambda'\}}\star g^i &=& \lambda'^if_{{\{\lambda\}\{\lambda'\}}}
\end{eqnarray}
is called the phasespace eigenfunction\cite{rf:1}  of $g^i$, 
and $\{\lambda\}$ ($\{\lambda'\}$)  is called the set of left (right) eigenvalues.
The above equations are sometimes called the stargen-value equations with respect to $g^i$. 
For  observables $O^i$ and an observable or a state function $\phi\in{\cal S}$,
let us define
\begin{eqnarray}
    {\phi}^{(O)}_{\{\lambda\}\{\lambda'\}} := \int \frac{dqdp}{(2\pi\hbar)^N}~{\phi}(q, p)\star O_{\{\lambda'\}\{\lambda\}}(q, p), \label{mtrix0}
\end{eqnarray}
where $O_{\{\lambda'\}\{\lambda\}}$ is an eigenfunction of $O^i$. 
We  call this quantity  the  matrix element of $\phi$ with respect to the observables $O^i$. 
Note the order of the subscripts in the above equation.

In terms of the above quantities, we make the following assumptions (3)-(5) 
concerning  measurements.

\noindent (3) An ideal measurement of a set of observables $O_i$ results in  a specific  set of values, 
and we assume each of these values to be one of left-right eigenvalues of $O_i$:
\begin{eqnarray}
  O_i\star O_{\{\lambda\}\{\lambda\}} = O_{\{\lambda\}\{\lambda\}}\star O_i = \lambda_iO_{\{\lambda\}\{\lambda\}}.
\end{eqnarray}
(4) If the state is described by $\Psi$, the probability of obtaining a particular set of values 
$\lambda$ as a result of the measurement of observables $O^i\in{\cal S}$ is
\begin{eqnarray}
        P(\Psi; O; \{\lambda\}) = \Psi^{(O)}_{\{\lambda\}\{\lambda\}}, \label{prob}
\end{eqnarray}
where the r.h.s. is defined by (\ref{mtrix0}). 
We assume that the state function $\Psi(q, p)$ and the 
eigenfunction $O_{\{\lambda\}\{\lambda\}}\in{\cal S}$ are normalized as
\begin{eqnarray}
          \int \frac{dqdp}{(2\pi\hbar)^N}~O_{\{\lambda\}\{\lambda'\}}(q, p) &=& \delta_{\{\lambda\}\{\lambda'\}} ,    \label{norm1} \\  
          \int \frac{dqdp}{(2\pi\hbar)^N}~{\Psi}(q, p) &=& 1, \label{norm2}
\end{eqnarray}    
where the r.h.s. of (\ref{norm1}) is the Kronecker delta or the delta function in a discrete or 
continuous spectrum.

\noindent (5) After the  measurement  of an observable, the  state reduces to the state described 
by a stargen-value function 
of the observable with equal left and right stargen-values. 
In general, this function is not a linear combination of left-right 
stargen functions of unobserved quantities.

For a non-ideal observation, in which the observed value lies in 
a certain range, 
the probability is the sum of the r.h.s. of Eq.~(\ref{prob}) over that range.
Thanks to the Hermiticity of the star product, the probability (\ref{prob}) is a real number.
Furthermore, it can be made positive definite by a proper choice of the star product.

Now, the physical state described by a state function $\Psi\in{\cal S}$ that  satisfies the condition
\begin{eqnarray}
       \Psi\star\Psi = \Psi  \label{pure}
\end{eqnarray}
is called a pure state.  A linear combination  of state functions describing pure states is another state function.
Because the probability (\ref{prob}) is linear with respect to the state function, a 
linear combination does {\it not} respect  `quantum mechanical' coherence, and for this reason,    
we call it an incoherent mixture of these states.
Presumably, there may exist other state functions that are neither  pure states nor 
 incoherent mixtures and have no counterparts in ordinary quantum mechanics. 
In the $D$-dimensional harmonic oscillator treated in Section 3, 
some solutions to the stargen-value equation  correspond to  neither pure states nor mixtures 
and  violate the associative law of the star product. 
Because of this violation these states do not describe physical states.

The time development of our abstract system is
determined by the following assumption.

\noindent (6) A physical state function $\Psi$ satisfies 
the  {Schr\"{o}dinger}-Moyal equation,
\begin{eqnarray}
      \frac{\partial}{\partial t} \Psi = -\{\Psi, H\}_M,  \label{SM}
\end{eqnarray}
where 
\begin{eqnarray}
       \{A, B\}_M := \frac{1}{i\hbar}(A\star B - B\star A)
\end{eqnarray}
is  the Moyal bracket of $A$ and $B$, and 
$H\in{\cal S}$ is called the Hamiltonian of the system.
The observables are assumed to be time independent.

These  assumptions define  the scheme of our star quantization.
The concrete  procedure  of obtaining the physical predictions is as follows. 

\noindent (i) Define a suitable star product operation in the function space of well-behaved functions of the phasespace, which may have some singularities.

\noindent (ii) Define the maximal set of observables, including the Hamiltonian, which are functions of the phasespace.

\noindent (iii) Solve the stargen-value equations for the maximal set of observables, which may have some 
singularities but should satisfy the normalization conditions.

\noindent (iv) In the function space spanned by  the solutions obtained 
in  (iii) and the functions of all the observables, 
determine the maximal function space, ${\cal S}$,   
in which any function satisfies the assumptions for the star product.

\noindent (v) A measurement  of an observable determines the 
initial state function through the assumption (5).

\noindent (vi) Solve the {Schr\"{o}dinger}-Moyal equation to  get the state 
function at a later time. 

\noindent (vii) The probability of 
getting a specific value of an observable at that time is obtained 
by Eq.~(\ref{prob}).

The above framework resembles that of the deformation quantization 
scheme in formal appearance. The difference between the two schemes 
lies in the definition of the physical states. 
In  both schemes, the physical spectrum is assumed 
to be determined  by the left-right stargen-values of observables 
consisting of a maximal commuting set.
The function describing the state after a 
measurement that consists of a subset of the maximal set of commuting observables
is a left-right stargen-value function of the observable 
that is actually measured.
In the deformation quantization scheme, however, this function 
is assumed to be a linear combination of the left-right stargen-value 
functions of {\it all}  observables, some of which may not have been measured.  
Contrastingly, our scheme requires that the  
function be a left-right stargen-value function of only {\it those} 
observables that are actually  measured.  
The function space of our scheme is larger than that of the 
deformation quantization, and our function space may 
contain functions possessing singularities, as long as they 
satisfy the normalization conditions.
In particular, we can consider  functions that are not 
(linear combinations of) the off-diagonal Wigner functions.  
Therefore it is not obvious whether the associativity of the star product 
applied to those functions holds (see below).

An example of the star product is that defined by Groenewold:\cite{rf:1,rf:8}
\begin{eqnarray}
   A\star B(q, p) &=& e^{\frac{i\hbar}{2}\sum_{i}\left(\frac{\partial}{\partial q_i}\frac{\partial}{\partial {p'_i}} - \frac{\partial}{\partial {q'_i}}\frac{\partial}{\partial p_i}\right)}A(q,p)B(q',p')\biggm|_{q=q', p=p'}.  \label{star}
\end{eqnarray}
In what follows we do not consider other choices for the star product, 
dealing only with (\ref{star}).

Although our framework is self-contained and requires no reference to  the Hilbert space 
of ordinary quantization, we review, for the sake of comparison, some relations between the 
Hilbert space and the function space of the phasespace. 
For an operator $\hat{A}$ of the Hilbert space a function of the phasespace is defined by 
the Weyl correspondence:\cite{rf:9}
\begin{eqnarray}
       A_W(q, p) &=& \int du~e^{-\frac{i}{\hbar}pu}\left<q+\frac{u}{2}\Biggm|\hat{A}\Biggm|q-\frac{u}{2}\right>.  \label{Weyl}
\end{eqnarray}
Conversely, we see
\begin{eqnarray}
       \langle q|\hat{A}|q'\rangle &=& \int\frac{dp}{(2\pi\hbar)^N}~e^{\frac{i}{\hbar}p(q - q')}A_W\left(\frac{q + q'}{2}, p\right). \label{invWeyl}
\end{eqnarray}
These relations, however, do not always define a one to one map between the operators of the Hilbert space 
and functions of the phasespace. 
For example, the integral (\ref{invWeyl}) for the function $A_W = (q + ip)^{-1}$ has 
a singularity along $q+q'=0$, which makes it impossible to define the corresponding 
operator in the Hilbert space.
(For the case of  systems with constraints, see Ref.\cite{rf:10}.)

The Weyl correspondence is used in deriving various relations between quantities 
in ordinary quantum mechanics and 
those in the Moyal scheme. For example, the phasespace eigenfunction of an observable $A$ is $(|n\rangle\langle m|)_W$, 
where $|n\rangle$ is the eigenvector of $\hat{A}$, 
and the Wigner function $\Psi(q, p)$ is defined  in terms of the state vector $|\psi\rangle$ as $\Psi(q, p)=(|\psi\rangle\langle \psi|)_W$.
The correspondence between operator multiplication and the star 
product is given by the relation
\begin{eqnarray}
         (\hat{A}\hat{B})_W = A_W\star B_W.  \label{AstarB}
\end{eqnarray}

Finally, we comment on the associativity of the star product.
The associative law of the star product would seem to  
be guaranteed by the 
corresponding operator relation of the quantum mechanics through the 
relation (\ref{AstarB}).
However, this is not always the case, as there might exist functions 
of the phasespace that have no operator counterparts.

\vspace{.5cm}


\section{Application to the harmonic oscillator}

The Schr\"{o}dinger equation for the one-dimensional harmonic oscillator described by 
the Hamiltonian $H_1 = (p^2+q^2)/2$ has solutions with positive energy eigenvalues    
expressed in terms of Hermite polynomials.
There also exist solutions with negative energies, but they cannot be normalized and 
therefore are excluded. 
The eigenfunctions $w_{nm} = (|n\rangle\langle m|)_W$ 
corresponding to solutions with positive energies, {\it i.e.}, those for which $n, m\ge 0$,  
were calculated long ago \cite{rf:8} and found to be expressed in terms of the associated Laguerre 
polynomials:
\begin{eqnarray}
 w_{nm} \propto a^{m-n}f_0(z)L^{(m-n)}_{n}(z),  \qquad  f_0(z) = 2e^{-z/2}, 
\end{eqnarray}
where $a = (q + ip)/\sqrt{2},  z=4H_1/\hbar$.
These functions are solutions to the  stargen-value equations 
\begin{eqnarray}
 H_1\star w_{nm} = E_nw_{nm},\qquad w_{nm}\star H_1 = E_mw_{nm},
\end{eqnarray}
with $E_n = (n+1/2)\hbar$.
(For calculations of the Wigner functions in various models, see Ref.\cite{rf:11}.)

Now, let us examine the solutions that are not necessarily off-diagonal Wigner functions but 
satisfy the stargen-value equations with normalization conditions.
For example, the functions $w_{nm}$ with $m<0$ satisfy the stargen-value 
equation.
In the one-dimensional case, however, the observed energies are the right {\it and} 
left eigenvalues of the stargen-value equations, and the values $m<0$ are excluded.
Contrastingly, in the $D$-dimensional case with $D\ge 2$, the total energy can be positive even if 
the sub-energies for some dimensions are negative.
This is a peculiar feature of our scheme, whose basic function space has no 
reference to a Hilbert space, which is absent in the case of the ordinary 
quantum mechanics and in the deformation quantization scheme.

The Hamiltonian of the $D$-dimensional harmonic oscillator is
\begin{eqnarray}
     H = \frac12\sum_{i=1}^{D}(p_i^2 + q_i^2).
\end{eqnarray}
The stargen-value equations for the Hamiltonian are $H\star f = f\star H = Ef$.
For later convenience, we introduce the creation and annihilation coordinates:
\begin{eqnarray}
     a_i = \frac{1}{\sqrt{2}}(q_i + ip_i), \quad
     a_i^* = \frac{1}{\sqrt{2}}(q_i - ip_i),\quad
     z_i = \frac{4}{\hbar}a_ia_i^*.
\end{eqnarray}
The star product is written in terms of these coordinates as
\begin{eqnarray}
   A\star B(a, a^*) &=& e^{\frac{\hbar}{2}\sum_i\left(\frac{\partial}{\partial a_i}\frac{\partial}{\partial a'^*_i} - \frac{\partial}{\partial {a^*_i}}\frac{\partial}{\partial a'_i}\right)}A(a,a^*)B(a',a'^*)\biggm|_{a=a', a^*=a'^*}.
\end{eqnarray}
The solutions to the stargen-value equations are the direct products of the one-dimensional 
solutions, and are expressed as
\begin{eqnarray}
&& \qquad  f(a_i, z_i) = \prod_i(a_i)^{k_i}e^{-z_i/2}F_i(z_i),\\
&&  E = \left(n + \frac{D}{2}\right)\hbar,  \qquad n = \sum_{i=1}^Dn_i, \qquad  \sum_{i=1}^Dk_i = 0,
\end{eqnarray}
where each $F_i(z_i)$ satisfies
\begin{eqnarray}
          z_iF_i'' + (k_i + 1 - z_i)F_i' + n_iF_i = 0. \label{sv3}
\end{eqnarray}
Here, the $n_i$ are the separation constants of the $D$-dimensional equation, and the $k_i$ are 
powers of Laurent expansions and are integers.

The quantity $n_i$ is the quantum number of the left stargen-value equation, 
$H_i\star f= (n_i+ 1/2)\hbar f$,  of the sub-Hamiltonian $H_i=(q_i^2+p_i^2)/2$ of 
the $i$-th dimension, and the quantity $n_i+k_i$ is the quantum number of the right stargen-value equation, $f\star H_i = (n_i+ k_i+1/2)\hbar f$. 
At this stage, the sub-energy of each dimension is not necessarily positive.

The admissible solutions, whose normalization integrals over the range with large values 
of the coordinates are not divergent are determined in the appendix, and we see that the 
solutions for each dimension are (apart from the normalization constant) given by 
\begin{eqnarray}
         W_{\nu\kappa}(a, z) = a^{\kappa}f_0(z)F_{\nu\kappa}(z), \qquad f_0(z) = 2e^{-z/2},
\end{eqnarray}
with integers $\nu$ and $\kappa$, where
\begin{eqnarray}
         F_{\nu\kappa}(z) = \left\{ \begin{array}{c}  L^{(\kappa)}_{\nu}(z) \qquad \qquad \qquad {\rm for}\quad \nu \ge 0,\\
              z^{-\kappa}L^{(-\kappa)}_{\nu+\kappa}(z) \qquad \qquad {\rm for}\quad 0 > \nu \ge -\kappa. \end{array} \right. \label{addmF} 
\end{eqnarray}
The energy eigenvalues of these solutions are 
\begin{eqnarray}
    E_{left} = \left(\nu + \frac12\right)\hbar,\qquad E_{right} = \left(\nu + \kappa + \frac12\right)\hbar.
\end{eqnarray}
Note that the solutions with 
negative $\nu$ (and $\nu+\kappa\ge 0$) have negative left energies, 
though they have positive right energies. 
(Here we consider stargen-value equations with sub-Hamiltonians.)

One possible argument for excluding such undesirable functions 
may be to show  
the divergence of the normalization integrals 
over the singularity  at the origin of the phasespace.
Since the functions considered here have different right and left eigenvalues of 
the sub-Hamiltonian, the normalization integrals should vanish.
Writing $a=re^{i\theta}$, the integrals are of the form 
$\int e^{i\kappa\theta}r^{1-\kappa}f(r)drd\theta$, where $f(r)$ is a 
regular function at the origin.
Although the integrals vanish if they are first integrated over $\theta$,
 they are not well-defined for $\kappa>0$, except in the case  $\kappa =1, \nu=-1$.
This fact shows that we cannot exclude at least the undesirable functions 
$W_{-1,1} = af_0F_{-1,1}$  by the requirement of the normalization condition.

Another possible argument for selecting admissible functions is the 
requirement of the associativity of the star product for such functions. 
It can be verified that the star product operation is well-defined on 
the function space spanned by positive powers of $a$ and $a^*$ and the 
solutions (\ref{addmF}) with $\nu\ge 0$.  It can also be shown that 
the associativity rule is satisfied.
Now, for the $\nu < 0$ case, we see that, for example, 
\begin{eqnarray}
       a^{*}\star W_{-1,1} = 2f_0, \qquad f_0\star a^{*} = 0, \qquad f_0\star f_0 = f_0,  \label{aWf}
\end{eqnarray}
where  $W_{-1,1}$ has negative left eigenvalues of the sub-Hamiltonian. From (\ref{aWf}) we see
\begin{eqnarray}
       (f_0\star a^{*})\star W_{-1,1} = 0, \qquad  f_0\star(a^{*}\star W_{-1,1}) = 2f_0,
\end{eqnarray}
{\it i.e.}, the violation of associativity.
The origin of this is seen from the inverse Weyl map, (\ref{invWeyl}), of $W_{-1,1}$, which diverges along $q+q'=0$. 
Therefore, the requirement of associativity excludes the undesirable solutions 
that have well-defined normalization integrals.
Similarly, other solutions  having negative energies are shown to violate associativity.
These functions cannot be written as linear combinations of functions satisfying (\ref{pure}).

The admissible observable functions are also determined by the requirement of 
associativity. For example, we see
\begin{eqnarray}
       (f_0\star a^{*})\star a^{*-1} = 0, \qquad  f_0\star(a^{*}\star a^{*-1}) = f_0.
\end{eqnarray}
Similarly, all negative powers of $a$ and $a^*$ are shown to violate associativity.

Thus the basic function space, ${\cal S}$, of the $D$-dimensional harmonic oscillator consists of 
the function space spanned by the positive energy solutions to the stargen-value 
equations, supplemented by the observable functions generated by 
the positive powers of $a_i$ and $a^{*}_i$.
The energy spectrum coincides with that of ordinary quantum mechanics.
This offers evidence suggesting the equivalence of the present scheme and quantum mechanics.

\newpage


\makeatletter
  \renewcommand{\theequation}{%
     A.\arabic{equation}}
  \@addtoreset{equation}{section}
\makeatother
\setcounter{equation}{0}

\noindent {\Large {\bf Appendix}}\vspace{.5cm}

In order to find the admissible solutions to Eq.~(\ref{sv3}), we study 
the solutions to the confluent hypergeometric equation in general. 
The parameters $n_i$ appearing there can be non-integers at this stage. 
The solutions to the confluent hypergeometric equation,
\begin{eqnarray}
z\frac{d^2u}{dz^2}+(\gamma-z)\frac{du}{dz}-\alpha u=0,
\end{eqnarray}
are written symbolically with the confluent $\tilde P$ function as
\begin{eqnarray}
u=\tilde P \left\{
		\matrix{\infty & 0 & {}          \cr
                  \overbrace{0 \quad \quad \alpha} & 0 & z \cr
                  ~~1  \quad \gamma-\alpha  & 1-\gamma & {}
                }
            \right\}.\label{cnflP}
\end{eqnarray}
This shows that the four explicit solutions are given by either functions 
that are analytic (except for poles) around $z=0$ or such functions 
around $z=\infty$, which are the singularities of the equation. 
At most two of them are independent.

The solutions around $z=0$ are given by $u_1=F(\alpha,\gamma;z)$ and $u_2=z^{1-\gamma}F(\alpha-\gamma+1,2-\gamma;z)$ for $\gamma\notin Z_+$, where $Z_+$ is the set of positive integers
(and, similarly, $Z_-$ represents the negative integers). 
(The case $\gamma\in Z_+$ will be discussed shortly.)
Here, the hypergeometric function of fluent type $F(\alpha,\gamma;z)$ is defined  by
\begin{eqnarray}
F(\alpha,\gamma;z)=\sum_{l=0}^{\infty}\frac{(\alpha)_n}{(\gamma)_n}\frac{z^n}{n!},
\end{eqnarray}
where $(\alpha)_n =  \alpha(\alpha+1)\cdots(\alpha+n-1)$.
We restrict ourselves to the case of integer $\gamma$ for the application of the present paper. 
The corresponding solutions constitute the basis of the Laurent expansions as mentioned 
in the main text.

Note that these solutions play the role of probability distribution 
functions when multiplied by $f_0=e^{-z/2}$. 
Therefore we should impose the condition that the solutions diverge more slowly 
than $e^{z/2}$ as $z\rightarrow\infty$, so that their integrations 
over phasespace are finite, which turns out to restrict the parameter region. 
A sufficient condition to realize this is the truncation of the power series of $z$, 
and it is in fact necessary also, because otherwise they diverge as $e^{z}$ in the 
limit $z\rightarrow\infty$.
This restricts the parameters such that either $\alpha\in Z_-\cup\{0\}$ or 
$\alpha - \gamma + 1\in Z_-\cup\{0\}$.
Introducing $n=-\alpha$ and $k=\gamma-1$ (in accordance with (\ref{sv3})), 
the solutions can be written in terms of the associated Laguerre 
polynomials, $L^{(k)}_n(z)$, as
\begin{eqnarray}
    u_1 &=& F(-n,k+1;z) = L_n^{(k)}(z), \qquad \qquad \qquad {\rm for}\quad n \ge 0,  \\
    u_2 &=& z^{-k}F(-(n+k),1-k;z)=z^{-k}L_{n+k}^{(-k)}(z), \quad {\rm for}\quad  n \ge -k, \label{sol_01}
\end{eqnarray}
where $k\le -1$.
These two solutions appear to be different, 
but they are in fact identical in the range $n \ge -k \ge 1$ 
which is seen by the identity
\begin{eqnarray}
z^{-k}L_{n+k}^{(-k)}(z)=(-1)^k\frac{n!}{(n+k)!}L_n^{(k)}(z),\qquad  {\rm for}\quad n, \quad n+k \ge 0.
\end{eqnarray}

In the case $\gamma\in Z_+$, the solutions are $\hat{u}_1=F(\alpha,\gamma;z)$ and $\hat{u}_2$, 
where 
$\hat{u}_2 = u_1\log{z} + z^{1-\gamma}F^*$ for  $\alpha\in Z_-\cup\{0\}\cup\{1\}$ or 
$\alpha - \gamma + 1\in Z_+\cup\{0\}$, and $\hat{u}_2 =z^{1-\gamma}F(\alpha-\gamma+1,2-\gamma;z)$ 
otherwise. 
Here $F^*$ is an analytic function around $z=0$.
The solution containing $F^*$ is excluded, since it diverges as $e^{z}$ 
in the limit $z\rightarrow\infty$. 
Therefore, the admissible solutions are
\begin{eqnarray}
        \hat{u}_1 &=& L_n^{(k)}(z), \qquad \qquad {\rm for}\quad n \ge 0, \\
        \hat{u}_2 &=& z^{-k}L_{n+k}^{(-k)}(z), \quad {\rm for}\quad 1 > n \ge -k,
\end{eqnarray}
where $k\ge 0$.

Other solutions represented by Eq.~(\ref{cnflP}) are 
those around $z=\infty$, which might include solutions 
having essential singularities at $z=0$. 
One of these diverges as $e^{z}$ in the limit $z\rightarrow\infty$, and 
therefore we exclude it. 
Another solution is obtained by setting $u=(1/z)^{\alpha}f(1/z)$. 
By solving the  equation for $f(1/z)$, 
we obtain the polynomial solutions obtained above.

We now summarize the solutions to Eq.~(\ref{sv3}). 
The solutions for $n_i \ge 0$ are the associated Laguerre polynomials, $L^{(k_i)}_{n_i}(z_i)$, 
and those for $0 > n_i \ge -k_i$ are $z^{-k_i}L^{(-k_i)}_{n+k}(z_i)$. 
Defining
\begin{eqnarray}
         \tilde{L}^{(k)}_{n}(z) = \left\{ \begin{array}{c}  L^{(k)}_{n}(z) \qquad \qquad \qquad {\rm for}\quad n \ge 0, ~~~~~~~{}\\
              z^{-k}L^{(-k)}_{n+k}(z) \qquad \qquad {\rm for}\quad 0 > n \ge -k, \end{array} \right. \label{Ltild}
\end{eqnarray}
the general solution of the left stargen-value equation, with total energy
\begin{eqnarray}
      E_{left} = \left(n + \frac{D}{2}\right)\hbar,  \label{leftE}
\end{eqnarray}
is written as a linear combination of functions of the form 
\begin{eqnarray}
    W(n_1,.., n_D;k_1,.., k_D) = f_0(z)(a_1)^{k_1}\cdot\cdot\cdot(a_D)^{k_D}\tilde{L}^{(k_1)}_{n_1}(z_1)\cdot\cdot\cdot\tilde{L}^{(k_D)}_{n_D}(z_D), \label{solution}
\end{eqnarray}
where $k_i$ are arbitrary integers  and  $n_i$ are integers greater than or equal to $-k_i$, satisfying 
$\sum_{i=1}^{D}n_i = n$. (For the case $D=1$ see Ref.\cite{rf:12}.)


\end{document}